\documentclass[aps,prb,twocolumn,floatfix,superscriptaddress]{revtex4-2}
\usepackage{graphicx}
\usepackage[english]{babel}
\usepackage{amsmath}
\usepackage{amssymb}
\usepackage{tensor}

\newcommand{\be}{\begin{eqnarray}}
\newcommand{\ee}{\end{eqnarray}}
\newcommand{\p}{\partial}

\newcommand{\dc}{c^{\dagger}}

\def\ep#1{\langle #1 \rangle}

\begin{document}

\title{Particle and thermal transport through one dimensional topological systems via Lindblad formalism}

\author{Yan He}
\affiliation{College of Physics, Sichuan University, Chengdu, Sichuan 610064, China}
\email{heyan$_$ctp@scu.edu.cn}

\author{Chih-Chun Chien}
\affiliation{Department of Physics, University of California, Merced, CA 95343, USA.}
\email{cchien5@ucmerced.edu}

\begin{abstract}
We apply the Lindblad quantum master equation to two examples of one-dimensional topological systems, the Su-Schrieffer-Heeger (SSH) model and Kitaev chain, to study their particle and thermal transport. The steady-state properties are obtained by decomposing fermions into Majorana fermions and extracting their correlation functions. We focus on the particle and thermal currents flowing through the bulk when the system is driven by two reservoirs coupled to the two ends. The ratio of the currents of the SSH model from the topological and trivial regimes with the same bandwidth demonstrates suppression of transport due to the edge states, which couple to the reservoirs but do not participate in transport. A similar comparison cannot be performed for the Kitaev chain because the topological and trivial regimes have different bandwidths, and the edge states are less significant away from the transition. Therefore, the results contrast different topological properties in quantum transport.
\end{abstract}

\maketitle

\section{Introduction}
Transport has been an important means for revealing signatures of topological materials. For example, the integer quantum Hall conductance is proportional to the Chern number~\cite{Niu1984}, and its generalization provides  early experimental evidence of topological insulator~\cite{BHZ-2}. Meanwhile, the Thouless pump has been realized in cold-atom systems~\cite{Nakajima2016,Lohse2016} to demonstrate quantized transport through modulation of topological systems, and Ref.~\cite{Citro22} summarizes recent progresses. There are reviews on topological transport in Dirac electronic systems ~\cite{Song17} and semimetals~\cite{Hu19}, showing the close connection between topology and transport.

On the theoretical side, transport coefficients may be extracted from linear response theory, Green's function methods, or other means~\cite{FetterBook,DiVentraBook,NazarovBook,StefanucciBook}. The Lindblad quantum master equation was designed to generate completely positive evolution of a system influenced by its environment~\cite{weiss2012quantum,Open-quantum-book} and is suitable for studying transport beyond the linear-response limit with a simplified description of the system-reservoir coupling. For example, the dependence of electronic conductance of a 1D hopping model on the system-reservoir coupling ~\cite{Gruss16} and effects from extended reservoirs have been discussed using normal fermions~\cite{Elenewski17} and the Kitaev chain \cite{DAbruzzo21} in the Lindblad formalism.  Effects of coupling strength, impurities, and disorder of fermionic systems in the Lindblad formalism have been analyzed in Refs.~\cite{PhysRevB.80.035110, PhysRevB.103.115139,PhysRevA.105.042214}. When applied to normal fermions with a quadratic Hamiltonian, the results from the Lindblad equation agree with those from the Landauer formalism~\cite{PhysRevE.94.032139,Lai18}. 
The Lindblad equation has been applied to particle and thermal transport in bosonic systems~\cite{Benenti_2009} and show geometry-based local circulation~\cite{Dugar20, Dugar22}. For bosonic systems with a quadratic Hamiltonian, the results from the Lindblad equation also agree with those from the quantum Langevin equation~\cite{Dugar22}. 

Here we use the Lindblad equation to investigate particle and thermal transport of two paradigmatic 1D topological systems, the SSH model~\cite{SSH79} and Kitaev chain~\cite{Kitaev-chain}. The former is an example of a topological insulator while the latter is a topological superconductor. The SSH model has been realized by cold-atoms in optical superlattices~\cite{Atala2013,Meier16,Kanungo22}, chlorine on Copper surface~\cite{Drost17}, graphene nanoribbons~\cite{Rizzo18,Groning18}, and electric circuits~\cite{LeeTopolectric18}. Some measurements of thermal transport of polymer-based materials may be related to the SSH model~\cite{Xu19}. On the other hand, the Kitaev chain may be realized by quantum dots~ \cite{Dvir22} or simulated on quantum computers~\cite{Stenger21,Huang21,Mi22}. There have been theoretical studies of the SSH model in the presence of disorder or long-range interactions~\cite{Klett18,PhysRevB.107.035113,Malakar23} and similarly for the Kitaev chain~\cite{Diehl11}. 

For the study of thermal transport, we note that differentiating the thermodynamic definitions of the thermal current related to the work and heat \cite{de2018reconciliation,hewgill2021quantum} helps resolve the concern on the thermodynamic consistency of the local (position basis) form of the Lindblad equation in the literature \cite{purkayastha2016out, cattaneo2019local, stockburger2017thermodynamic, levy2014local}. For example, there are multiple ways of defining the thermal current in open quantum systems \cite{asadian2013heat,hewgill2021quantum, prosen2010exact}. Reference \cite{asadian2013heat} presents two different expressions of the thermal current of a linear chain, one derived at the boundary and one derived in the bulk. As explained in Ref. \cite{hewgill2021quantum}, the expression at the boundary contains both heat and work while the one in the bulk is directly from heat. Here we will focus on the thermal current associated with heat in our discussion of thermal transport.

Our results will show a topological suppression of particle and thermal transport of the SSH model due to the localized edge states. In contrast, topological properties of the Kitaev chain do not have significant influence of particle and thermal transport within the Lindblad formalism. We will show the difference comes from the dependence of their topological properties on the bandwidth and the form of their edge states. We mention that particle transport of the SSH model has been studied using the Green's function method~\cite{Pineda22}, showing edge-state signatures in the local density of state.  Meanwhile, the Kitaev chain has been studied using the Green's function method~\cite{Leumer21}, quantum Langevin equation plus Green's function~\cite{Bhat20}, and Green's function plus master equation~\cite{Jin22} to show maps of the conductance and by linear response theory~\cite{Quan22} to analyze thermoelectricity. Studies of thermal transport through classical harmonic systems described by classical Langevin equations have shown suppression of the thermal conductance due to the emergence of localized edge modes~\cite{Chien18} and geometry-based circulation~\cite{Dugar19}. Moreover, localized edges states have been observed in heat diffusion through mechanical systems~\cite{Qi22}. Our exact results will clarify the correspondence between specific topological properties and transport phenomena via the two concrete examples.

The rest of the paper is organized as follows. Sec.~\ref{sec:Linblad} outlines the Lindblad formalism and its application to the study of quantum transport in the two exemplary topological systems. Sec.~\ref{sec:current} shows the correlations functions that lead to the particle and thermal currents. Sec.~\ref{sec:result} presents the suppression of transport due to the edge states in the SSH model. Transport in the Kitaev chain, in contrast, is shown to be dominated by the bandwidth and insensitive to topological effect. Sec.~\ref{sec:conclusion} concludes our study. Some derivations are summarized in the Appendix. 

\section{Lindblad formalism of quantum transport}\label{sec:Linblad}
To obtain the expressions of the particle and thermal currents through a 1D quantum system of size $N$, we assume that the two ends of the system are coupled to two different fermionic baths with different temperatures. Here we will consider a local version of the Lindblad master equation with the dissipator operators only applied to the sites at the two ends of the system. The equation of motion of the density matrix of the system is given by
\be
\frac{d\rho}{dt}
&=&-i[H,\rho]+\sum_{\nu}\Big(L_{\nu}\rho L_{\nu}^{\dag}
-\frac12\{L_{\nu}^{\dag}L_{\nu},\rho\}\Big) \nonumber \\
&=&-i[H,\rho]\nonumber+\Big[\gamma_L N_L\Big(\dc_1\rho c_1-\frac12\{c_1 \dc_1,\rho\}\Big)+\nonumber \\
&&\gamma_L (1-N_L)\Big(c_1\rho \dc_1-\frac12\{\dc_1 c_1,\rho\}\Big)\Big]+\nonumber\\
&&\Big[\gamma_R N_R\Big(\dc_N\rho c_N-\frac12\{c_N \dc_N,\rho\}\Big)+\nonumber \\
&&\gamma_R (1-N_R)\Big(c_N\rho \dc_N-\frac12\{\dc_N c_N,\rho\}\Big)\Big].
\label{eq-rho}
\ee
Here $\gamma_L$ and $\gamma_R$ are the system-reservoir coupling constants. $N_L$ and $N_R$ are fermion numbers in the baths. Here we assume a single-mode approximation of the baths, so they only exchange particles with the system at the chemical potential $\mu$, leading to
\be\label{eq-Nres}
N_s=\frac{1}{\exp(-\mu/T_s)+1},\quad s=L,\,R.
\ee
If we choose $T_L>T_R$, it will drive a thermal current along the chain from left to right. The matching of the energy levels in the reservoirs with $\mu$ is to avoid additional energy loss when the fermions are exchanged between the system and reservoirs, as it is known that energy mismatches at the system-reservoir interfaces can lead to energy dissipation in the Landauer treatment of ideal conductors~\cite{DiVentraBook}. Eq.~\eqref{eq-Nres} thus allows us to focus on the effects from different temperatures of the reservoirs. We caution that $\mu$ is non-vanishing in the single-mode approximation, or there will be no transport because the two reservoirs will have the same particle density regardless of their temperatures. One may go beyond the single-mode approximation and have different values of $\mu_{L,R}$ of the reservoirs and $\mu$ for the system to investigate additional energy dissipation, which is beyond the scope of the present work. We remark that it is straightforward to generalize the set up to have the reservoirs connected to multiple sites of the system.

We will contrast the quantum transport of particles and heat through two 1D topological systems using the Lindblad quantum master equation. Our first example is the SSH model with the Hamiltonian~\cite{SSH79,Asboth2016}
\be\label{eq:HSSH}
H&=&\sum_{m=1}^{N/2}\Big[w_1(\dc_{2m-1} c_{2m}+\dc_{2m}c_{2m-1})+ \nonumber \\
&& w_2(\dc_{2m} c_{2m+1}+\dc_{2m+1}c_{2m})\Big]
-\sum_{j=1}^N\mu\dc_j c_j.
\ee
Here $c_j$ and $c_j^{\dagger}$ are the fermion annihilation and creation operators on site $j$, respectively, and $N$ is the total number of sites. The topology of the SSH model is characterized by the winding number with periodic boundary condition or the localized edge states with open boundary conditions~\cite{Asboth2016}, and $w_2/w_1 > 1$ (or $<1$) corresponds to the topological (or trivial) regime. 
Due to the coupling to the reservoirs, an onsite potential playing the role of the chemical potential $\mu$ is also introduced. The presence of $\mu$ breaks the chiral symmetry of the SSH model, but the winding number and edge states are not affected. 

The second example is the Kitaev chain~\cite{Kitaev-chain}. The SSH model represents a topological insulator while the Kitaev chain is a p-wave topological superconductor. The Hamiltonian of the Kitaev chain in real space is given by
\be\label{eq:HK}
H&=&\sum_{j=1}^N\Big[-w(\dc_j c_{j+1}+\dc_{j+1}c_j)-\mu\dc_j c_j+\nonumber \\
&&\Delta(\dc_j\dc_{j+1}+c_{j+1}c_j)\Big].
\ee
Here $\Delta$ is the order parameter from nearest-neighbor pairing. The topology of the Kitaev chain is characterized by the $\mathbb{Z}_2$ index of the Majorana number with periodic boundary condition and the emergence of localized edge states with open boundary condition~\cite{Kitaev-chain}. For our studies of quantum transport, we assume open boundary condition for the above two models. We remark that the SSH model respects particle conservation in isolation while the Kitaev chain does not. Ref.~\cite{DAbruzzo21} discussed implications of particle non-conserving effects of the Kitaev chain with extended reservoirs.

\subsection{Time evolution of correlation functions}
For fermionic systems with quadratic Hamiltonians, it is more challenging to directly solve the density matrix from Eq. (\ref{eq-rho}) than finding the time evolution of the correlation functions and then obtaining the physical quantities of interest from the correlation functions. To this end, it is instructive to define the Majorana fermion operators as
\be\label{eq:Majorana}
c_j=\frac12(a_{2j-1}+ia_{2j}),\quad  \dc_j=\frac12(a_{2j-1}-ia_{2j}),
\ee
which satisfy $\{a_i,a_j\}=2\delta_{i,j}$. Then we define the following correlation functions in terms of the Majorana fermions as
\be\label{eq:Kmn}
K_{nm}=\textrm{Tr}(\rho\hat{\Gamma}_{nm}),\quad \hat{\Gamma}_{nm}=\frac i2[a_n,a_m].
\ee
Here the commutator is anti-symmetric, so a factor $i$ is included to make the operator $\hat{\Gamma}_{nm}$ Hermitian.

To find the time evolution of $K_{mn}$, we use the method outlined in Ref. \cite{Bardyn2013} to transform the Lindblad equation \eqref{eq-rho}
with a quadratic Hamiltonian and linear dissipator operators to the equation of motion for the correlations. $H$ and $L_{\mu}$ are expressed in term of the Majorana fermions as
\be\label{eq:HLL}
H=\sum_{jk}\frac {i h_{jk}}2 a_j a_k,\quad L_\mu=\sum_j l_{\mu j}a_j,\quad
L_\mu=\sum_j l_{\mu j}^*a_j.
\ee
Here $h_{ij}$ denote the elements of a real anti-symmetric matrix $h$ satisfying $h_{ij}=-h_{ji}$. We find that the equation of motion of the correlation matrix $K$ is given by
\be
\frac{\p}{\p t}K=2[h,K]-2\{X,K\}-4Y.
\label{K-eq}
\ee
Here $X$ and $Y$ are the real and imaginary parts of the matrix $M$ defined as
\be
M_{jk}=\sum_\mu l^*_{\mu j}l_{\mu k},\quad M=X+i Y.
\ee
Here $M$ is a Hermitian matrix. Then $X$ and $Y$, as the real and imaginary parts of $M$, must satisfy the relations $X=X^T$ and $Y=-Y^T$. The derivation of Eq. (\ref{K-eq}) is shown in Appendix \ref{sec-app-K}.

Now we apply the general formalism to the Lindblad equation (\ref{eq-rho}). It can be shown that $M$ is a block-diagonal matrix given by
\be
&&M=\textrm{diag}\Big\{M_1,M_2,\cdots,M_N\Big\}.
\ee
Here $M_1=\frac{\gamma_L}4[\sigma_0+(1-2N_L)\sigma_2]$ and
$M_N=\frac{\gamma_R}4[\sigma_0+(1-2N_R)\sigma_2]$, where $\sigma_0$ is the 2 by 2 identity matrix, and $\sigma_i$ with $i=1,2,3$ are the Pauli matrices. Only the first and last $2\times2$ blocks along the diagonal are non-zero. Then the real matrix $X$ is given by
\be
&&X=\textrm{diag}\Big\{X_1,X_2,\cdots,X_N\Big\},
\ee
where $X_1=\frac{\gamma_L}4\sigma_0$ and
$X_N=\frac{\gamma_R}4\sigma_0$.
Similarly, $Y$ is given by
\be
&&Y=\textrm{diag}\Big\{Y_1,Y_2,\cdots,Y_N\Big\},
\ee
where $Y_1=\frac{\gamma_L}4(1-2N_L)(-i\sigma_2)$ and
$Y_N=\frac{\gamma_R}4(1-2N_R)(-i\sigma_2)$.
Again, only the first and last $2\times2$ blocks along the diagonal are non-zero.

After setting up the Lindblad equation and the corresponding equations for the correlations, we rewrite the Hamiltonian in terms of the Majorana fermions.
For the SSH model,
\be
H&=&\frac{i}2\Big[\sum_j(-\mu a_{2j-1}a_{2j})+\sum_{j\in odd}w_1(a_{2j}a_{2j+1}+\nonumber\\ &&a_{2j-1}a_{2j+2})+\sum_{j\in even}w_2(a_{2j}a_{2j+1}+a_{2j-1}a_{2j+2})\Big]. \nonumber \\
&&
\ee
The nonzero elements of $h$ are given as follows.
\be
h_{2j-1,2j}&=&-h_{2j,2j-1}=-\frac{\mu}2,\\
h_{2j,2j+1}&=&-h_{2j+1,2j}=-h_{2j-1,2j+2}=h_{2j+2,2j-1} \nonumber \\
&=&-\frac{w_1}2,\quad j\in odd;\\
h_{2j,2j+1}&=&-h_{2j+1,2j}=-h_{2j-1,2j+2}=h_{2j+2,2j-1} \nonumber \\
&=&-\frac{w_2}2,\quad j\in even.
\ee
For the Kitaev chain,
\be
H&=&\frac{i}2\Big[\sum_{j=1}^{N-1}\Big(-\mu a_{2j-1}a_{2j}+(\Delta+w) a_{2j}a_{2j+1}+\nonumber \\
& &(\Delta-w) a_{2j-1}a_{2j+2}\Big)-\mu a_{2N-1}a_{2N}\Big].
\ee
The nonzero elements of $h$ are given as follows.
\be
&&h_{2j-1,2j}=-h_{2j,2j-1}=-\frac{\mu}2,\\
&&h_{2j,2j+1}=-h_{2j+1,2j}=\frac{\Delta+w}2,\\
&&h_{2j-1,2j+2}=-h_{2j+2,2j-1}=\frac{\Delta-w}2.
\ee

\subsection{Steady state}
For a non-zero $Y$, there will be at least one steady state. In the cases studied here, we only found one steady state for each setup. The correlation functions in the steady state $K_s$ satisfy
$[h,K_s]-\{X,K_s\}=2Y$.
It is convenient to rewrite the expression as
$(h-X)K_s-K_s(h+X)=2Y$,
which can be expressed by a system of linear equations \cite{Prosen-JSM} as
\be\label{eq:SS}
\Big[(h-X)\otimes I-I\otimes(h+X)^T\Big](K_s)_{\text{vec}}=2(Y)_{\text{vec}}.
\ee
Here $I$ is the identity matrix with the same dimension as $h$, and $Y_{\text{vec}}$ is the matrix $Y$ written as a vector.
Therefore, we are able to obtain the steady-state solution without the distraction of the transient behavior.


\section{Correlation functions and  currents}\label{sec:current}
The correlation functions of the original fermions can be expressed in terms of the Majorana-fermion correlation functions $K_{ij}$ as follows. For $i\neq j$,
\be
\ep{\dc_ic_j}&=&\frac14\Big[(K_{2i-1,2j}-K_{2i,2j-1})-\nonumber \\
&&i(K_{2i-1,2j-1}+K_{2i,2j})\Big],\\
\ep{c_ic_j}&=&\frac14\Big[(K_{2i-1,2j}+K_{2i,2j-1})-\nonumber \\
&& i(K_{2i-1,2j-1}-K_{2i,2j})\Big],
\ee
and for $i=j$, we find that
\be
\ep{\dc_ic_i}=\frac12\Big[1+K_{2i-1,2i}\Big].
\ee
Moreover, $\ep{c_ic_i}=0$, which is consistent with the Fermi statistics. 
From these correlation functions, we can compute many observable quantities. An example is the density at each site given by
\be
n_j=\ep{\dc_j c_j}.
\ee

\subsection{Particle and thermal currents of SSH model}
In order to define the particle and thermal currents of the SSH model, we write the Hamiltonian~\eqref{eq:HSSH} as $H=\sum_{j=1}^{N-1} H_{i,j+1}$ with
\be
H_{j,j+1}=\bar{w}_j(\dc_j c_{j+1}+\dc_{j+1}c_j)-\mu\dc_j c_j.
\ee
Here we define $\bar{w}_j$, given by
$$
\bar{w}_j=\left\{
      \begin{array}{ll}
        w_1, & j\in\text{odd}, \\
        w_2, & j\in\text{even}.
      \end{array}
    \right.
$$
Then the particle current operator from site $j$ to site $j+1$ can be defined as
\be
(\hat{J}_p)_{j,j+1}=\frac{d n_j}{dt}\Big|_R\equiv i[\dc_j c_j, H_{j,j+1}].
\ee
Here $|_R$ means that the time change is only caused by the Hamiltonian across the sites $j$ and $j+1$.
With this definition, the particle current operator is
\be
(\hat{J}_p)_{j,j+1}=i \bar{w}_j(\dc_j c_{j+1}-\dc_{j+1}c_j).
\ee

In order to define the thermal current, we consider the Hamiltonian of the left part of system. To be precise, the left part includes the lattice sites form  $i=1$ to $i=j$ with $1<j<N$. Thus,
\be
H_L=\sum_{i=1}^j H_{i,i+1}.
\ee
The thermal current operator between the $j$ and $j+1$ sites is then given by
\be
(\hat{J}_t)_{j,j+1}=\frac{d H_L}{d t}=i[H_L, H]=i[H_{j,j+1},H_{j+1,j+2}].
\ee
A straightforward calculation of the above commutator gives
\be
(\hat{J}_t)_{j,j+1}&=&i\Big[w_1w_2(\dc_j c_{j+2}-\dc_{j+2}c_j)-\nonumber \\
&&\bar{w}_j\mu(\dc_j c_{j+1}-\dc_{j+1}c_j)\Big].
\ee
The particle and thermal currents of the SSH model, $J_{p,t}=Tr(\rho \hat{J}_{p,t})$ can be expressed in terms of the fermion correlation functions. We have a remark about the particle and thermal currents in the bulk: The operator forms of the currents seem to have no information regarding the reservoirs or system-reservoir coupling. This is because in the Lindblad formalism, the system density matrix already contains those information. The bulk-current operators, on the other hand, are local operators within the system. When taking the ensemble average, the expectation values $J_{p,t}$ reflect the total properties of the systems, reservoirs, and system-reservoir coupling. We also caution that the bulk-current expressions should not be used at the system-reservoir interfaces ($j=1$ or $N$) because there may be additional contributions of thermodynamic work in the expressions, as explained in Ref.~\cite{hewgill2021quantum}. 

\subsection{Particle and thermal currents of Kitaev chain}
The same definitions also apply to the Kitaev chain with the Hamiltonian~\eqref{eq:HK} expressed as $H=\sum_{j=1}^{N-1} H_{i,j+1}$, where
\be
H_{j,j+1}&=&-w(\dc_j c_{j+1}+\dc_{j+1}c_j)-\mu\dc_j c_j+\nonumber \\
&&\Delta(\dc_j\dc_{j+1}+c_{j+1}c_j).
\ee
Then the particle current operator from site $j$ to site $j+1$ is given by
\be
(\hat{J}_p)_{j,j+1}=-i w(\dc_j c_{j+1}-\dc_{j+1}c_j)+i\Delta(\dc_j \dc_{j+1}-c_{j+1}c_j).
\label{eq-Jp-K}
\ee
The first and second parts represent the contributions from single-particles and pairs, respectively. 

In order to find the thermal current operator of the Kitaev chain, we have to calculate the following commutator
\be
(\hat{J}_t)_{j,j+1}=i[H_{j,j+1},H_{j+1,j+2}].
\ee
To simplify the notation, we take $j=1$, and define $H_{12}=A+B+C$, with $A,B,C$ denoting the hopping, density, and pairing terms.
Similarly, we also write $H_{23}=A'+B'+C'$. Then it follows $[B,A']=[B,B']=[B,C']=0$. The non-zero commutators are given by
\be
&&[A,A']=w^2(\dc_1c_3-\dc_3c_1),~
[A,B']=w\mu(\dc_1c_2-\dc_2c_1)\nonumber\\
&&[A,C']=-w\Delta(\dc_1\dc_3-c_3c_1),\nonumber \\
&&[C,A']=-w\Delta(-\dc_1\dc_3+c_3c_1),\nonumber\\
&&[C,B']=-\mu\Delta(-\dc_1\dc_2+c_2c_1),\nonumber \\
&&[C,C']=\Delta^2(-\dc_1c_3+\dc_3c_1).
\ee
Collecting the above results, we find
\be
(\hat{J}_t)_{j,j+1}&=&i\Big[(w^2-\Delta^2)(\dc_j c_{j+2}-\dc_{j+2}c_j)+\nonumber \\
&&w\mu(\dc_j c_{j+1}-\dc_{j+1}c_j)
+\mu\Delta(\dc_j\dc_{j+1}-c_{j+1}c_j)\Big].\nonumber \\
\label{eq-Jt-K}
\ee
Similar to the SSH model, both particle and thermal currents of the Kitaev chain can also be computed from the fermion correlation functions.

\section{Results and discussions}\label{sec:result}
Here we present the numerical results of the thermal and particle currents of the SSH model and Kitaev chain and explore the influence from topology. We will mainly focus on the properties in the steady state in the long-time limit, which is more meaningful as the initial-state information decays away. Here we focus on the setup with two baths coupled to the first and last sites of the system. The results presented here are taken with symmetric system-reservoir couplings: $\gamma_L=\gamma_R=\gamma$. We have verified that using reasonable asymmetric system-reservoir couplings only introduces quantitative changes.

\begin{figure}
\centering
\includegraphics[width=\columnwidth]{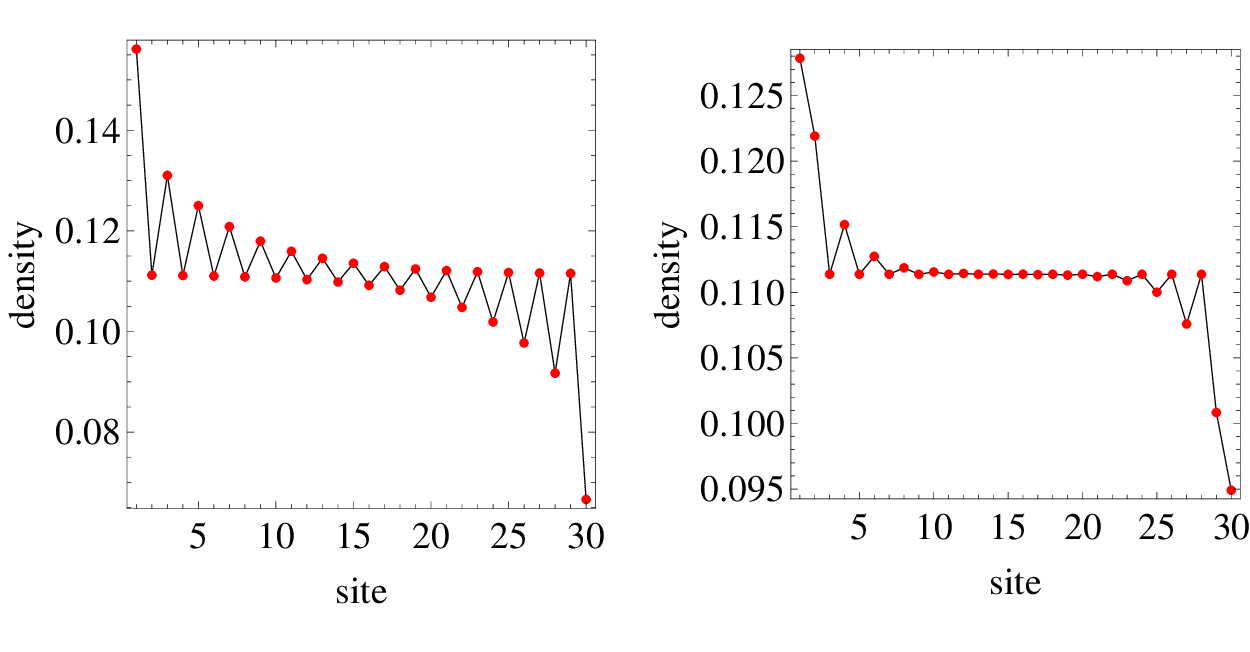}
\caption{Steady-state density profiles of the SSH model. The first and last sites are connected to two reservoirs with $T_L=0.8$ and $T_R=0.1$. The left (right) panel is topological (trivial) with $w_1=1$ and $w_2=1.2$ ($w_1=2$ and $w_2=1.2$), respectively. Here $\mu=1$, $\gamma=1$, and $N=30$. }
\label{topo}
\end{figure}

\subsection{SSH model}
We first discuss the results of the SSH model with $\mu$ set to be the unit, i.e., $\mu=1$. The system-reservoir coupling constants are chosen to be symmetric and set to $\gamma_L=\gamma_R=1$. Tuning $\gamma_{L,R}$ only leads to quantitative changes. The temperatures of the baths are $T_L=0.8$ and $T_R=0.1$, and we have checked that choosing different temperatures does not qualitatively change the conclusions. However, this does not exclude the possibilities of interesting phenomena when $N_{L,R}$ are small in more complicated settings. We first consider the topological case with $w_1=1$, $w_2=1.2$ and set the total number of lattice sites to $N=30$. Since the steady-state thermal and particle currents in the system are all flat due to particle and energy conservation, their plots do not reveal much information. In contrast,
Figure \ref{topo} shows the steady-state density profile from the solution of Eq.~\eqref{eq:SS} for selected topological and trivial cases. 
The oscillations with the densities on alternating site roughly equal is a signature of the localized states in the SSH model in the topological regime with the bulk states forming a relatively flat background. In contrast, the oscillations in the density profile in the trivial regime decay rapidly with the system size. 

To verify the steady-state results from Eq.~\eqref{eq:SS}, we also integrate the Lindblad equation~\eqref{eq-rho} numerically and confirm the long-time steady-state results agree with the exact solution from the steady-state correlation functions. We remark that while the SSH model is a topological insulator with a band gap, the presence of the reservoirs and system-reservoir coupling leads to particle exchanges throughout the chain. This is different from the insulator picture in the thermodynamic limit, where the lower band is fully occupied and the reservoirs are only allow to exchange particles within the lower band, which will be prohibited by the Pauli exclusion principle.

\begin{figure}
\centering
\includegraphics[width=\columnwidth]{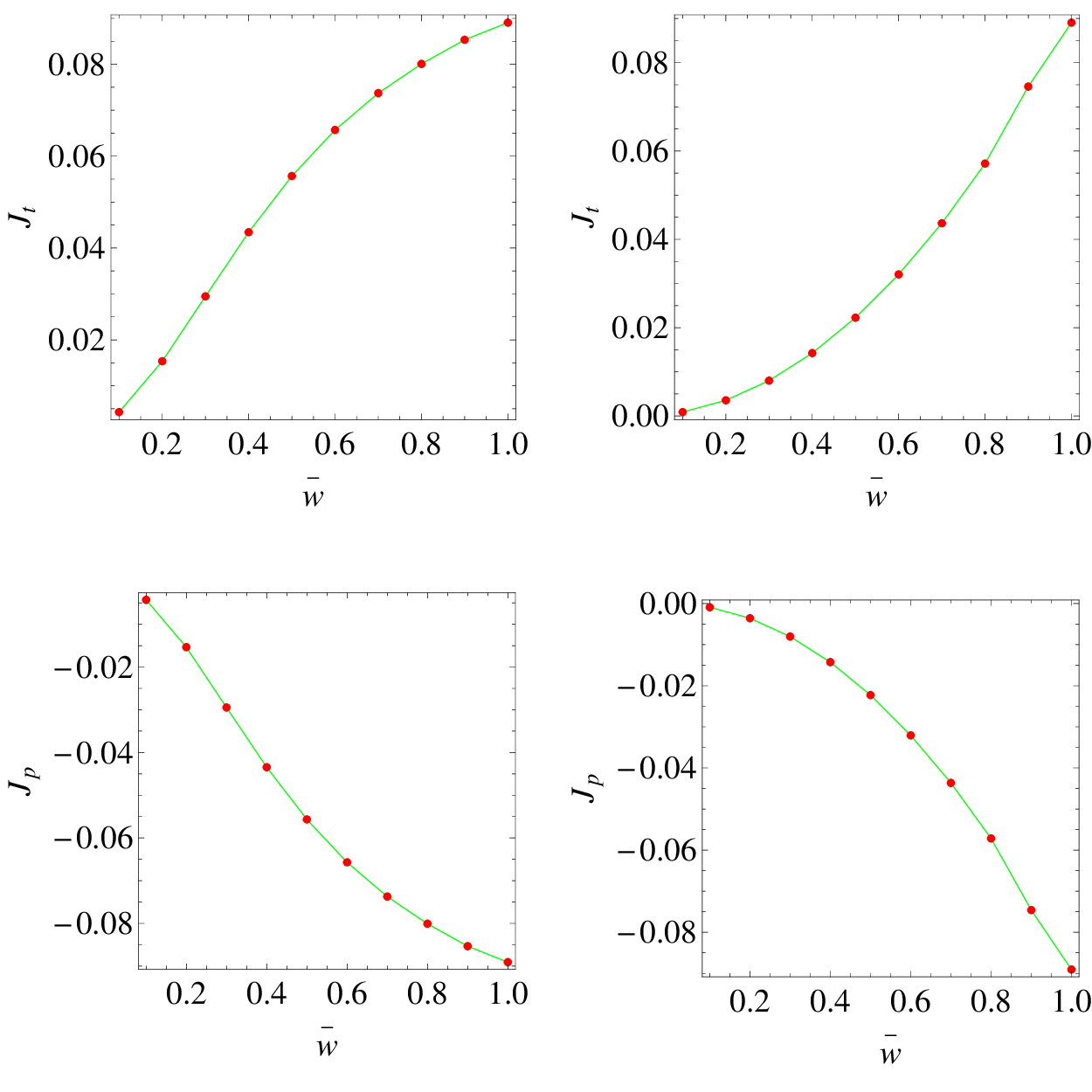}
\caption{Steady-state currents in the middle of the SSH model. (Upper left) Thermal current $J_t$ as a function of $\bar{w}=w_2<1$ with $w_1=1$ in the topologically trivial regime. (Upper right) Thermal current $J_t$ as a function of $\bar{w}=w_1<1$ with $w_2=1$ in the topological regime. (Bottom row) The corresponding particle currents $J_p$. Here $\gamma_{L,R}=1$, $T_L=0.8$, $T_R=0.1$, and $\mu=1$.}
\label{Jt-SSH}
\end{figure}

The thermal current has the opposite sign of the particle current in this case. Actually, the thermal and particle currents of the SSH model are proportional to each other. Since the SSH model only involves the nearest neighbor hopping and the baths only couple to the fist and last sites, one may expect the only nonzero correlations occur between nearest-neighboring sites. This is indeed the case for the SSH model because numerically we find that Im$\ep{\dc_j c_{j+2}}=0$. Therefore,
\be
J_t=-i\bar{w}\mu\ep{\dc_j c_{j+1}-\dc_{j+1}c_j}=-\mu J_p.
\ee
Since we choose a positive value of $\mu$, the thermal current $J_t$ is opposite to the particle current $J_p$. The relation shows that the thermal and particle currents have opposite signs if $\mu >0$ but the same sign if $\mu<0$. This is because higher $T$ and positive $\mu$ leads to a lower density in the bath according to the Fermi statistics~\eqref{eq-Nres} that attracts particles but sends out heat, causing the opposite signs between the thermal and particle currents.

To understand how the band topology affects the thermal and particle currents, we will focus on the exact solution of the current through the system in the steady state. 
In the upper left panel of Figure \ref{Jt-SSH}, we plot the steady-state thermal current at the center of system as a function of $w_2$ with fixed $w_1=1$. As $w_2$ decreases from $1$ to zero, the inter-cell hopping is always smaller than the intra-cell hopping, and the system stays in the non-topological regime. One can see the thermal current decreases with $w_2$ as the bandwidth shrinks. The curve is slightly curved upward. Similarly, in the upper right panel of Figure \ref{Jt-SSH}, we fix $w_2=1$ and decrease $w_1$ from $1$ to zero. In this case, the inter-cell hopping is always larger than the intra-cell hopping, and the system stays in the topological regime. The thermal current $J_t$ also decreases with $w_1$ but slightly curved downward, implying that it decreases faster than the non-topological case. 

\begin{figure}
\centering
\includegraphics[width=\columnwidth]{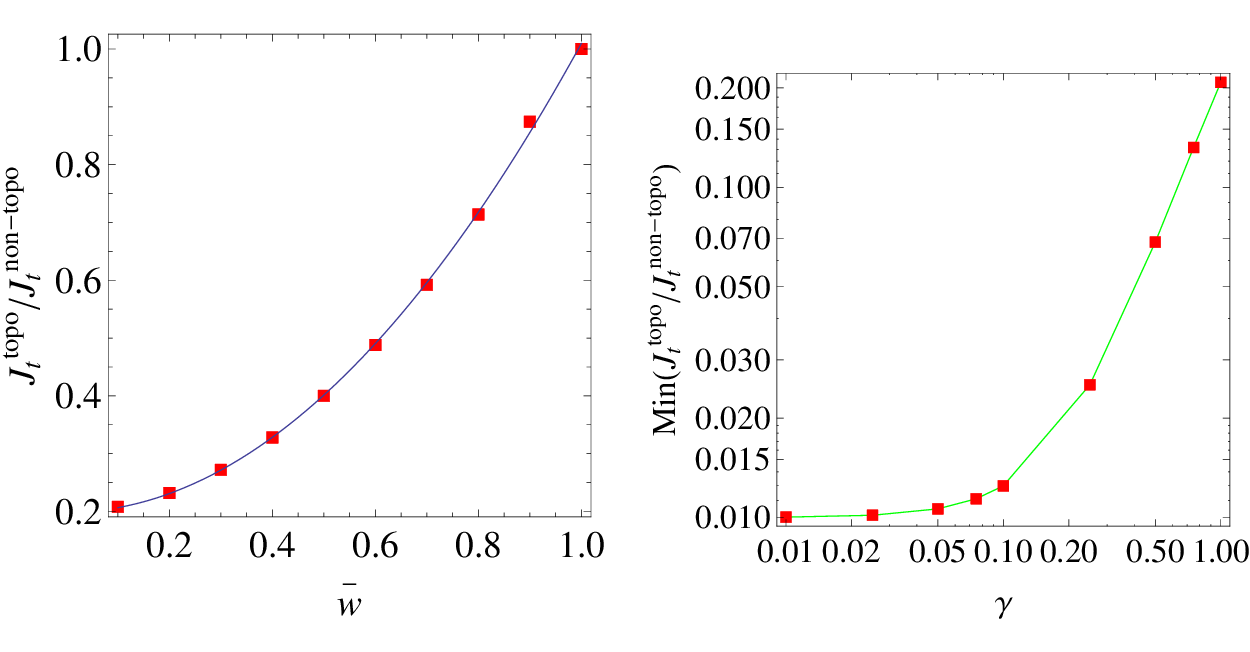}
\caption{(Left) Ratio of the steady-state thermal currents of the SSH model from the topological and trivial cases shown in Fig. \ref{Jt-SSH}. The curve is a fit to a quadratic form with a constant term. The ratio of the steady-state particle currents exhibits the same behavior (see the context). (Right) The constant term (or the minimum) from the fit of the ratio as a function of $\gamma$.}
\label{J1J2}
\end{figure}

To extract topological effects from the results, we will use the notation $\bar{w}=min(w_1, w_2)$ when $max(w_1, w_2)=1$.  We remark that by fixing one of the hopping coefficient and decreasing the other one, the bandwidths decrease in the same manner in the topological and non-topological cases. The ratio between the thermal or particle current with the same $\bar{w}$ thus allows for a fair comparison to extract the effect from the band topology because the bandwidths are the same.
In Figure \ref{J1J2}, we plot the ratio of the thermal currents in the middle of the system from the topological and non-topological cases as a function of $\bar{w}$. Since the particle current is proportional to the thermal current for the SSH model, the ratio for the particle current is the same as that of the thermal current. 

Although the currents in both topological and non-topological cases approach zero as $\bar{w}$ vanishes, the ratio between the currents shows that the topological case suppresses the current more significantly as $\bar{w}$ decreases. 
The ratio in Figure \ref{J1J2} can be fitted to a quadratic function of the form 
\be
J_t^{\text{topo}}/J_t^{\text{non-topo}}\approx C_1 + C_2 \bar{w}^2.
\ee
We found that the quadratic dependence is insensitive to the value of $\gamma$. However, the term $C_1$, which is the minimum of the ratio as $\bar{w}\rightarrow 0$ increases with $\gamma$, and its dependence is shown in the right panel of Figure \ref{J1J2}. As explained below, the edge states are responsible for the quadratic suppression of the ratio of 
the thermal currents as the hopping coefficients change. Meanwhile, the system-reservoir coupling shifts the ratio by adding a $\gamma$-dependent background to the quadratic term.

The quadratic dependence of the ratio of currents from the topological and non-topological cases on $\bar{w}$ may be understood by analyzing the edge states. With open boundary condition, the topological case has a pair of edge states, one localized on the left and the other localized on the right, while the non-topological case only has delocalized states. By diagonalizing the SSH Hamiltonian, the left edge state has the amplitude $v_0=\mathcal{N}(1, 0, -\bar{w}, 0, (-\bar{w})^2, \cdots)^T$ along a long chain. The normalization condition $|v_0|^2=1$ gives $\mathcal{N}^2=1-\bar{w}^2$. The right edge state is arranged oppositely. If we consider the unitary matrix $U=(v_0, v_{B1}, v_{B2},\cdots)$ by collecting the eigenstates of the Hamiltonian, including the localized edge states and the bulk states forming the bands, the unitarity requires that each column and each row should be normalized to $1$. The first row, however, depicts the weights of the states at the left end of the chain. Therefore, the normalization along the first row shows that the bulk states only has a total weight of $1-\mathcal{N}^2=\bar{w}^2$ if an edge state emerges on the left. However, the edge state decays exponentially and does not participate in transport. The edge state on the right end also contributes the same suppression of weights. Thus, the effective transport bandwidth of the topological case is reduced to a fraction $\bar{w}^2$ relative to the non-topological case. Our numerical results catch the correct power-law dependence of the suppression from the edge states already at intermediate size of the chain. Finally, we remark that the inclusion of the chemical potential in the SSH model breaks the chiral symmetry, but it does not change  the winding number and the edge states. 

\begin{figure}
\centering
\includegraphics[width=0.9\columnwidth]{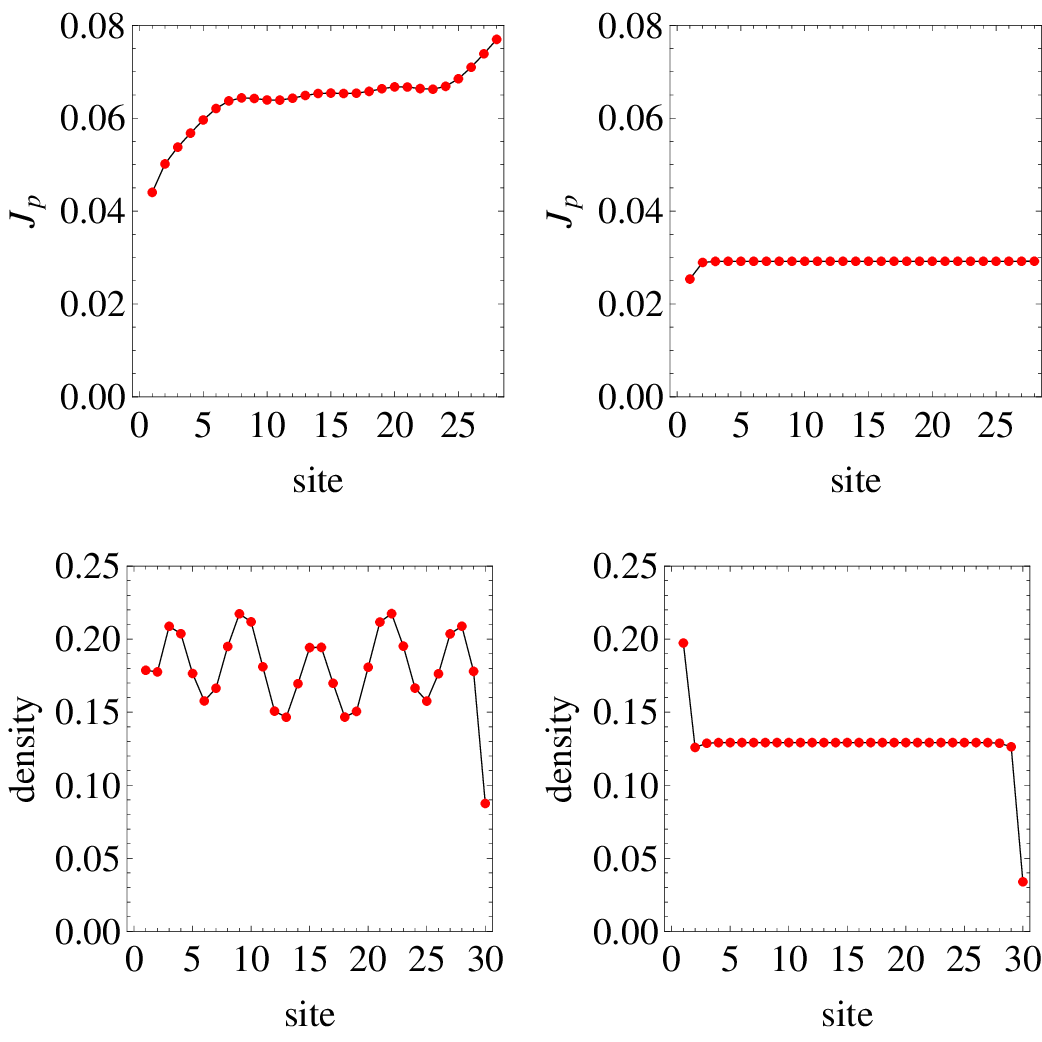}
\caption{Steady-state profiles of the local particle current $J_p$ (top row) and density (bottom row) of the Kitaev chain.  The left (right) column is topological (trivial) with $w=0.6$ ($w=0.3$). The first and last sites are connected to two reservoirs with $T_L=0.8$ and $T_R=0.1$. Here $\mu=1$, $\Delta=0.1$, $\gamma=1$ and $N=30$.
}
\label{K-topo}
\end{figure}



\subsection{Kitaev chain}
Next, we investigate the particle and thermal currents of the Kitaev chain. The topology of the SSH model is determined by the ratio between $w_1$ and $w_2$. For the Kitaev chain with finite $\Delta$, the topology is determined by the ratio between $2w$ and $\mu$ when $|\Delta| >0$. The topological (or trivial) region corresponds to $w/\mu > 1/2$ (or $w/\mu < 1/2$). The profiles of the thermal current $J_t$ is flat in both topological and trivial regimes, as expected from energy conservation. In contrast, the particle current and density profiles distinguish the topological and trivial regimes. Figure \ref{K-topo} displays the real-space profiles of the particle current $J_p$ and density of in the steady state of the Kitaev chain in the topological and trivial regimes. While $J_p$ remains flat in the trivial regime, there are observable deviations from the bulk value near the reservoirs in the topological regime. Similarly, the density shows some small oscillations in the topological regime while it is flat in the trivial regime. However, we found that the density oscillations in the topological regime slowly decrease as the system size increases, so the oscillations are finite-size effects. Meanwhile, the deviations of $J_p$ at the two ends do not scale with the system size, so we will focus on the flat bulk current in the middle of the system.  
Moreover, we have verified that numerical integration of the Lindblad equation~\eqref{eq-rho} agrees with the steady-state result from Eq.~\eqref{eq:SS} in the long-time limit.

\begin{figure}
\centering
\includegraphics[width=\columnwidth]{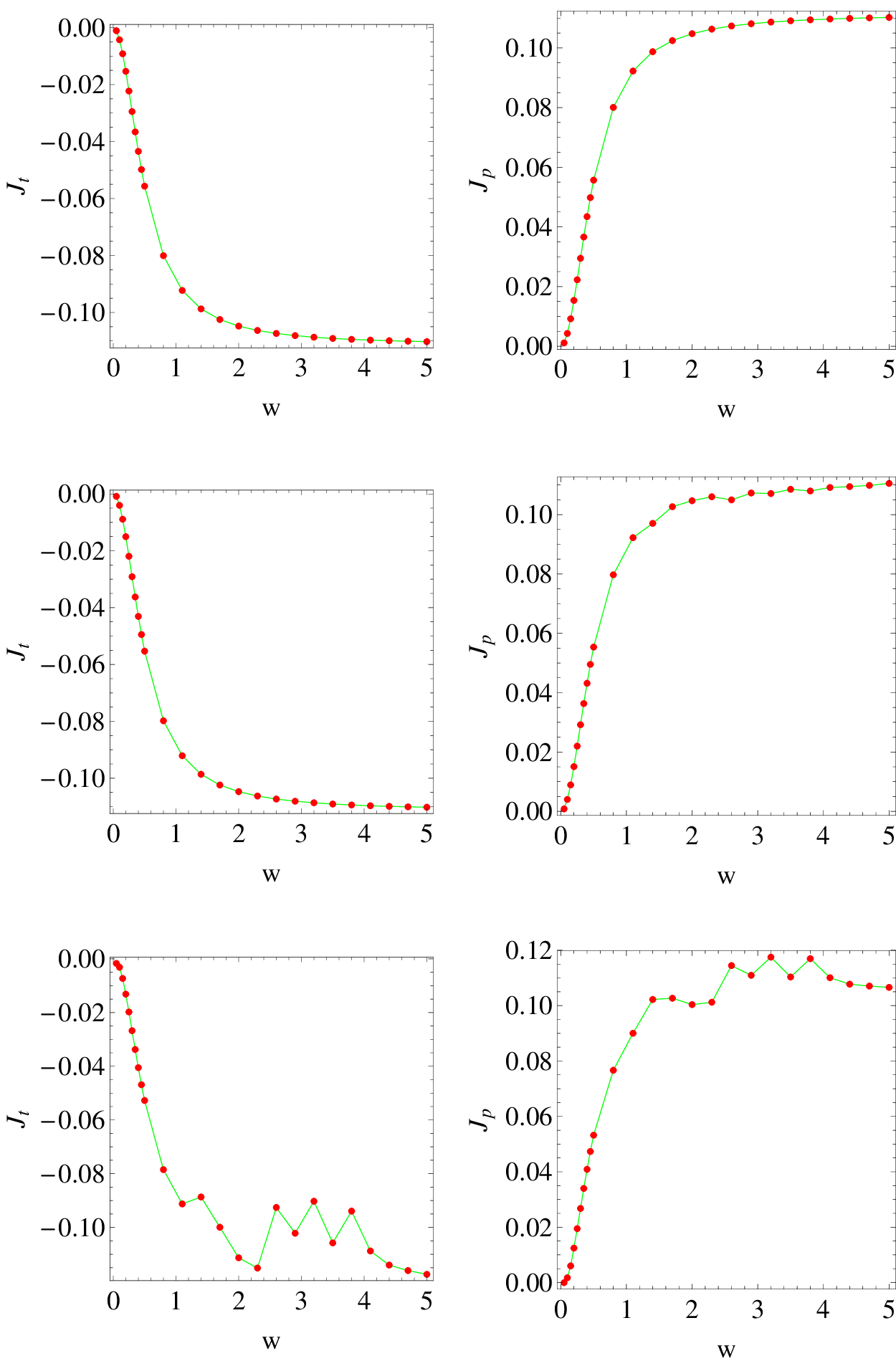}
\caption{Steady-state thermal current $J_t$ (left column) and particle current $J_p$ (right column) in the middle of the Kitaev chain versus $w$ with $\mu=1$ and $\Delta=0,~ 0.1,~ 0.3$ (from top to bottom). The topological (trivial) region corresponds to $w>1/2$ ($w<1/2$). Here $\gamma=1$, $T_L=0.8$, $T_R=0.1$, and $N=30$.}
\label{fig:K0}
\end{figure}

To see the dependence of transport on the bandwidth controlled by $w$, we first set $\Delta =0$ and find the steady-state particle and thermal currents in the middle of the system. The results are shown in Fig~\ref{fig:K0}. While the magnitude of the currents increases linearly with $w$ initially, the currents saturate in the large-$w$ limit. The saturation of the particle current may be understood by the quantum of conductance in a ballistic channel~\cite{DiVentra2009,NazarovBook}. For the Kitaev chain, we can find analytical expressions of the steady-state correlation function $K_s$ and the particle current $J_p$ in the large-$w$ limit. Explicitly, 
\be\label{eq:KJp}
J_p\rightarrow\Big(\frac1{\gamma_1\gamma_2}+\frac{1}{4w^2}\Big)^{-1}\frac{N_R-N_L}{\gamma_1+\gamma_2}.
\ee
A sketch of the derivation is given in Appendix~\ref{app:largew}. We note that the particle current only depends on the external parameters in the infinite-$w$ limit. 

When $\Delta$ is added to the Hamiltonian, the steady-state thermal and particle currents in the middle of the system qualitatively follow the $\Delta=0$ results, as shown in Fig.~\ref{fig:K0}. There is no drastic changes around the critical point $w=(1/2)\mu$. As $\Delta$ increases, the oscillatory behavior when $w$ becomes large and makes it challenging to extract the functional dependence. We found the amplitudes of the oscillations in the particle or thermal current of the large-$\Delta$ cases decrease slowly as the system size increases, indicating that those oscillations are finite-size effects and the curve should be smooth in the thermodynamic limit.

According to Eq. (\ref{eq-Jt-K}), the thermal current of the Kitaev model has three terms. In our study, the contributions from the next-nearest neighbor hopping term and the pairing term are negligible compared to the nearest neighbor hopping term. By comparing the thermal current with the particle current of Eq. (\ref{eq-Jp-K}) and noting that $\mu=1$, we found that $J_t\approx-J_p$ for the Kitaev chain. Thus, we can focus on the $J_p$ curve.
Meanwhile, the analysis in the large-$w$ limit summarized in Appendix~\ref{app:largew} applies to the case with finite $\Delta$ as well. Therefore, the particle current will approach a constant as $w\rightarrow\infty$. Our numerical results suggest the thermal current behaves qualitatively the same as the particle current when $\Delta$ is finite. 

Different from the results of the SSH model, the steady-state currents of the Kitaev chain from the Lindblad formalism with the single-mode approximation of the reservoirs do not reveal discernible influence from its topological properties.  There are two reasons behind the difference. Firstly, the SSH allows a change of its topological indicator without a change of its bandwidth by swapping the alternating hopping coefficients. In contrast, the topology of the Kitaev chain changes with the bandwidth if the chemical potential is fixed. Therefore, the effects from topological change is buried in the more dominant effect from increasing bandwidth. The quantum-channel limit of transport in the large-$w$ regime further constrains the influence from band topology. Secondly, the localization of the edge states of the SSH model become more prominent away from the critical point $w_2/w_1=1$ because the amplitude decays from the edge with powers of $w_1/w_2$. In contrast, the localization of the edge states of the Kitaev chain becomes less prominent away from the critical point $w=(1/2)\mu$ because the amplitude decays from the edge with powers of $\sqrt{w^2-\Delta^2}/(w+\Delta)$~\cite{He2016a}. We remark that topological effects in transport through the Kitaev chain may stand out in more complex treatments beyond this study.

In the bottom row of Figure \ref{fig:K0}, we plot the steady-state thermal current $J_t$ and particle $J_p$ at the center of the chain as a function of $w$ with $\mu=1$ and $\Delta=0.3$. For larger pairing gap, the currents display stronger oscillations in the real space. This makes the curves of $J_t$ and $J_p$ more bumpy than the small-$\Delta$ cases, but they still show the same qualitative behavior as previous discussed. Therefore, the transport through the Kitaev model from the Lindblad equation is dominated by the bandwidth of the corresponding noninteracting system, leaving no significant trace of topological effects.

\subsection{Implications}
We emphasize that taking the ratio of the particle or thermal currents between the topological and trivial cases of the SSH model leads to a fair comparison because the topological criterion can be implemented without changing the bandwidth. Therefore, the ratio of the currents is controlled by the edge states and system-reservoir coupling. In contrast, the topological properties of the Kitaev chain vary with the bandwidth, hindering a fair comparison between transport in the topological and trivial regimes. While Ref.~\cite{Bhat20} allows the chemical potentials of the reservoirs to differ and scans the chemical potentials to map the conductance, here we treat the chemical potential as a constant to focus on how topological properties affect thermal or particle transport. Moreover, we have verified that the results are insensitive to the system size once the number of lattice sites becomes reasonably large.

For the two systems studied here, we only include the particle exchange operators in the Lindblad equation. While this is reasonable for the SSH model, pairing effects are only included in the Hamiltonian of the Kitaev chain as an effective model. The insignificance of topological effects in transport through the Kitaev chain is thus at the single-particle level. 
We also remark that the derivation of the Lindblad equation implicitly limits the system-reservoir coupling to the weakly interacting regime~\cite{weiss2012quantum,Open-quantum-book}. Therefore, the results with $\gamma$ larger than the system bandwidth may not be realistic when compared to experimental results, as strong-interaction effects are expected to modify the approximation. One has to use the full quantum master equation or other means~\cite{weiss2012quantum,Open-quantum-book} to describe the quantum dynamics with strong system-reservoir coupling. Our results also show that identifying prominent topological signatures in transport is already a challenging task in simple setups before considering disorder or long-range interaction that will further enrich the nonequilibrium physics of topological systems.

\section{conclusion}\label{sec:conclusion}
We have shown that the exact solutions of the steady states of the Lindblad equation allow a detailed analysis of the particle and thermal transport through two paradigms in 1D topological systems, the SSH model and Kitaev chain. The results contrast how topological effects influence quantum dynamics: The ratio of the particle or thermal currents from the topological and trivial regimes of the SSH model with the same bandwidth reveals the suppression from the edge states due to their localized nature. Such an extraction of topological effects cannot be achieved in the Kitaev chain, as the transport is dominated by the different bandwidths in the topological and trivial regimes. While rapid advance in quantum materials, devices, and simulators may verify the results and test the limits of the Lindblad formalism, the framework and analysis may also be generalized to interacting systems in the future for studying the interplay between topology, interaction, and transport.

\begin{acknowledgments}
Y. H. was supported by the Natural Science Foundation of China under Grant No. 11874272 and Science Specialty Program of Sichuan University under Grant No. 2020SCUNL210. C. C. C. was supported by the National Science Foundation under Grant No. PHY-2011360.
\end{acknowledgments}

\appendix

\section{Derivation of the time-evolution equation in a matrix form}
\label{sec-app-K}
Here we derive the matrix form of the time-evolution equation (\ref{K-eq}) from the Lindblad equation given by the first line of Eq.~\eqref{eq-rho}.
The Hamiltonian and jump operators $L_\mu$ are shown in Eq.~\eqref{eq:HLL}.
We have defined the Majorana fermion operators in Eq.~\eqref{eq:Majorana}.
The correlation function is defined in Eq.~\eqref{eq:Kmn}.
Making use of the Lindblad equation, we find that
\be\label{eq:dKdt}
&&\p_t K_{nm}=\textrm{Tr}\Big(\p_t\rho\,\hat{\Gamma}_{nm}\Big)=\textrm{Tr}\Big(\rho[A+B]\Big).
\ee
Here $A=\sum_\mu\Big[L^{\dag}_{\mu}\hat{\Gamma}_{nm} L_{\mu}-\frac12\{L_{\mu}^{\dag}L_{\mu},\hat{\Gamma}_{nm}\}\Big], \label{K-A}$ and
$B=-i[\hat{\Gamma}_{nm},H]\label{K-B}$.
It is convenient to introduce the following quantities
\be
M_{jk}&=&\sum_\mu l^*_{\mu j}l_{\mu k},\quad \hat{\Gamma}_{nm}=\frac i2\sum_{jk}G^{nm}_{jk}c_j c_k,\nonumber \\
G^{nm}_{jk}&=&\delta_{nj}\delta_{mk}-\delta_{nk}\delta_{mj}.
\ee
Note that $M^*_{jk}=M_{kj}$, thus $M$ is Hermitian. 
After some algebra, we obtain
\be
A=\frac i4 M_{ij}G^{nm}_{kl}\Big(2a_i a_k a_l a_j-a_i a_j a_k a_l-a_k a_l a_i a_j\Big).
\ee
From the canonical commutation relations, we find that
\be
a_i a_k a_l a_j-a_i a_j a_k a_l=2(a_i a_k\delta_{jl}-a_i a_l\delta_{jk}),\\
a_i a_k a_l a_j-a_k a_l a_i a_j=2(a_l a_j\delta_{ik}-a_k a_j\delta_{il}).
\ee
Also making use of $G^{nm}_{kl}=-G^{nm}_{lk}$, we can rewrite $A$ as
\be
A&=&-i\Big(M_{ij}G^{nm}_{jk}a_i a_k+G^{nm}_{ki}M_{ij}a_k a_j\Big) \\
&=&-i a_j\Big(\{M,G^{nm}\}\Big)_{jk}a_k. \nonumber 
\ee
After decomposing $M$ into the real and imaginary parts as
\be
M=X+i Y,\quad X=X^T,\quad Y=-Y^T,
\ee
we arrive at an expression for $A$ as
\be
A=-i\Big(X_{nj}[c_j ,c_m]+[c_n,c_k]X_{km}-4iY_{nm}\Big).
\ee
Similarly, we can also rewrite $B$ as
\be
&&B=-i[\hat{\Gamma}_{nm},H]=\frac i2[c_n c_m,\,h_{jk}c_j c_k]\nonumber\\
&&=i\Big(h_{nk}[c_k,c_m]-[c_n,c_k]h_{km}\Big).
\ee
After plugging the above two equations for $A$ and $B$ into Eq.~\eqref{eq:dKdt}, we finally arrive at
the matrix form of the time-evolution equation (\ref{K-eq}) in the main text.

\section{Analytic expressions of the Kitaev chain in the large-$w$ limit}\label{app:largew}
Here we analytically solve the steady-state Lindblad equation of the Kitaev chain with $\Delta=0$. For small and finite $\Delta$, the result is also valid in the large-$w$ limit. We present the simplified case with $\gamma_L=\gamma_R=1$ first. It is more convenient to write $K_s$ as a $N\times N$ matrix consisting of $2\times 2$ blocks.
Inspired by the numerical results, we assume $K_s$ takes the following form
\be
K_s=\left(
      \begin{array}{ccccc}
        A_1 & B &  &  &  \\
        -B  & A & B &  &  \\
         & \ddots & \ddots & \ddots &  \\
         &  & -B & A & B \\
         &  &  & -B  & A_2
      \end{array}
    \right),
\ee
where the non-zero blocks are only along the diagonal, super- and sub-diagonal lines. The blocks $A$, $B$ and $A_{1,2}$ are assumed to have the following forms
\be
A&=&a(i\sigma_2),\quad A_1=(a+c)(i\sigma_2),\quad A_2=(a-c)(i\sigma_2),\nonumber\\ 
B&=&b \sigma_0,
\ee
where $\sigma_0$ is the 2 by 2 identity matrix, $\sigma_2$ is the second Pauli matrix, and $a,b,c$ are unknown numbers to be determined.

With this form of $K_s$, it is straightforward to compute the left hand side of the steady state equation as
\be
(h-X)K_s-K_s(h+X)=\left(
     \begin{array}{ccccc}
       C_1 & D &  &  &  \\
       -D & 0 &  &  &  \\
        &  & \ddots & 0 & D \\
        &  &  & -D & C_2
     \end{array}
   \right)
\ee
with $C_{1,2}$ and $D$ given by
\be
&&C_1=\left(
    \begin{array}{cc}
      0 & -\frac{a+c}2+bw \\
      \frac{a+c}2-bw & 0
    \end{array}
  \right),\nonumber \\
&&C_2=\left(
    \begin{array}{cc}
      0 & -\frac{a-c}2-bw \\
      \frac{a-c}2+bw & 0
    \end{array}
  \right),\nonumber\\
&&D=\frac14\left(
    \begin{array}{cc}
    -b-2cw & 0 \\
      0 & -b-2cw
    \end{array}
  \right).
\ee
Note that these are the only non-zero blocks. All the omitted blocks are zero. Moreover, those blocks do not depend on $\mu$ even if $h$ contains non-zero $\mu$. Thus the resulting $K_s$ will be independent of $\mu$. The right hand side of the steady state equation is given by
\be
2Y=\left(
     \begin{array}{ccccc}
       F_1 & 0& & & \\
       0 & 0 & & & \\
        & & \ddots & & \\
        & & & 0& 0 \\
        & & & 0 & F_2
     \end{array}
   \right).
\ee
The only non-zero blocks are $F_1=\frac12(2N_L-1)(i\sigma_2)$ and $F_2=\frac12(2N_R-1)(i\sigma_2)$. Equating the two sides, we find that $C_1=F_1$, $C_2=F_2$ and $D=0$.
Those conditions turn into the following algebraic equations:
$-b-cw=0$, 
$b w-\frac{a+c}{2}=\frac12(2N_L-1)$, and
$-b w-\frac{a-c}{2}=\frac12(2N_R-1)$.
The solution is
\be
&&a=-(N_L+N_R-1),\\
&&b=(1+\frac{1}{4w^2})^{-1}\frac{N_L-N_R}{2w}\\
&&c=-\frac b{2w}.
\ee
Then we find the expectation values of the hopping term and particle current as
\be
&&\ep{\dc_j c_{j+1}}=-\frac i4\Big[(K_s)_{2j-1,2j+1}+(K_s)_{2j,2j+2}\Big]=-\frac i2 b.\nonumber \\
\ee
When $\Delta=0$ or when the system is in the $w>>\Delta$ limit where the pairing term can be ignored, we find the particle current $(J_p)_{j,j+1}=2w\text{Im}\ep{\dc_j c_{j+1}}=-wb$ given by
\be
(J_p)_{j,j+1}&=&(1+\frac{1}{4w^2})^{-1}\frac{N_R-N_L}{2}\approx\frac{N_R-N_L}{2}.
\ee

For arbitrary values of $\gamma_{1,2}$, the calculations are similar but more tedious, we only show the expression of the particle current in Eq.~\eqref{eq:KJp}.
Importantly, the analytic expressions have been verified by the numerical results in the large-$w$ regime.

\bibliographystyle{apsrev}

\end{document}